\documentclass[conference]{IEEEtran}
\IEEEoverridecommandlockouts
\usepackage{url}
\usepackage{cite}
\usepackage{amsmath,amssymb,amsfonts}
\usepackage{algorithmic}
\usepackage{graphicx}
\usepackage{textcomp}
\usepackage{xcolor}
\def\BibTeX{{\rm B\kern-.05em{\sc i\kern-.025em b}\kern-.08em
    T\kern-.1667em\lower.7ex\hbox{E}\kern-.125emX}}
\begin{document}

\title{Rethinking the Memory Staleness Problem in Dynamics GNN\\
}

\author{\IEEEauthorblockN{Mor Ventura, Hadas Ben Atya, Dekel Brav}
\IEEEauthorblockA{\textit{MSc} \\
\textit{Technion - IIT}\\
 Haifa, Israel \\
\{mor.ventura, hds, dekel.brav\}@campus.technion.ac.il}
}

\maketitle

\begin{abstract}
The staleness problem is a well-known problem when working with dynamic data, due to the absence of events for a long time. Since the memory of the node is updated only when the node is involved in an event, the i's memory becomes stale. Usually it refer to lack of events such as a temporal deactivating of a social account \cite{TGN}. To overcome the memory staleness problem \cite{TGN} aggregate information from the node's neighbors memory in addition to the node's memory. Inspired by that, we design an updated embedding module that inserts the most similar node in addition to the node's neighbors.
Our method achieved similar results to the TGN, with a slight improvement.
This could indicate a potential improvement after fine-tuning our hyper-parameters, especially the time threshold, and using a learnable similarity metric.
\end{abstract}


\section{Introduction}
Many fields have turned to Graph Neural Networks (GNNs) for modeling systems of relations and interactions, such as social sciences and biology. While most of the models assume that the underlying graph is static, real-life systems of interactions are dynamic \cite{TGN}. 

Recently dynamic graphs have been researched and can be divided into two groups, discrete-time dynamic graphs (DTDG) and continuous-time dynamic graphs (CTDG).
However, discrete-time dynamic graphs are unsuitable for several real-world settings such as social networks in which the dynamic graph are continuous and evolving. 

On the other hand, continuous-time dynamic graph my suffer from a memory staleness problem, due to a lack of events. 

To handle the staleness of representations,  \cite{KumarLearningNetworks} propose a method to learn how the representation for a node i evolves when no observation involving the node is made \cite{staleness}. 

The TGN model \cite{TGN} mitigates the staleness problem by aggregating information from a node’s temporal neighbor's memory. According to their article, when a node has been inactive for a while, it is likely that some of its temporal neighbors have been recently active, and by aggregating their memories, TGN can compute an up-to-date embedding for the node. They used temporal graph attention to select which neighbors are more important based on both features and timing information.

We propose a new method, for nodes that were stale for relative a long period $\Delta T$, using the most similar nodes' information as well as the temporal neighbors.

Our code is available on GitHub \url{https://github.com/venturamor/dynamic_graph.git}.

\section{Related Work}
To represent accurately dynamic systems using graphs, there is a must to represent the temporal evolution within them. While most of the works focus on discrete-time dynamic graphs \cite{dynamicgraph1, dynamicgraphs2}, It has also been suggested to represent the continuous-time changes using a sequence of time-stamped graph nodes' embeddings \cite{TGN}. 
One of the main problems is the memory staleness problem.\cite{Kazemi2020RepresentationSurvey} When the update of node embedding is based on its memory, and it was not updated for a while due to lack of usage, for example when a user stops using a social network platform it causes stale memory, it may lead to worse performance, especially when the end task is recommendation system based on the edges.
The solution that was suggested in the baseline paper to mitigate this problem is to apply a TGN, graph embedding module with a temporal attention mechanism that is updated by aggregating information from a node's neighbors' memory \cite{TGN}. It is based on the assumption that though it, i.e the user, was not active recently, its neighbors were. This temporal graph attention takes into account the last N temporal neighbors. For instance, in the Wikipedia dataset, a temporal neighbor is one item of the 10 recent items that were edited by this user. Solutions that lack the consideration of a node's neighbors when computing its embedding are susceptible to the staleness problem.

\section{Datasets}
We used two datasets in our experiments, Wikipedia and Reddit. The Reddit dataset is a graph dataset from Reddit posts made in the month of September 2014. The nodes are the community, or “subreddit”, to which a post belongs to. An interaction occurs when a user writes a post to the subreddit\cite{RedditCode, TGN}. In the Wikipedia dataset, users and items (pages) are the nodes, and an interaction represents a user editing an item.
In both aforementioned datasets, the interactions are represented by text features, and labels represent whether a user is banned. Both interactions and labels are time-stamped, and each event, user, and item has a unique id \cite{TGN}.

\section{Method}
"Tell me who your friends are, and I will tell you who you are". This is the guiding saying of our method. We look for the most similar source nodes (users), based on their memory features and use theirs embedding when computing the current relative staled node.

\begin{figure}
    \centering
    \resizebox{0.95\columnwidth}{!}{\includegraphics{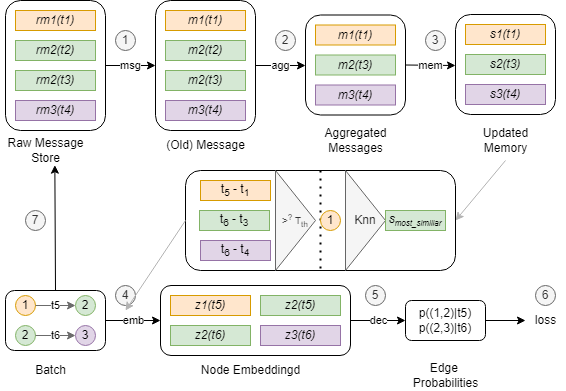}}
    \caption{The updated flow of operations of TGN. The embedding is computed through staleness check based on relative time difference and KNN algorithm to compute similarity to other nodes memory. (based on Figure 2 in \cite{TGN})}
    \label{fig:flow}
\end{figure}

Our method differentiate between short-time stale nodes and long-time stale. 
When a node was stale for a period time longer than our threshold, $\Delta T > \Delta T_{th}$, We update the node's embedding with the embedding from it's $ K $ most similar nodes, in additional to it's $ N $ temporal neighbors.

The new node's embedding can be written as the following equation: 
\begin{equation}
\resizebox{.8\hsize}{!}{$Emd(S_i) = Emd(S_i) + \sum_{k=1}^{K} Emb(S_k) +  GAT(\{S_{1}^{i},.., S_{N}^{i}\}) $}
\end{equation}

Where $Emd(S_i)$ is the i's node embedding representation, $Emb(S_k)$ is the embedding of the k'th most similar node to i, and $GAT(\{S_{1}^{i}, .., S_{N}^{i}\})$ is the Temporal Graph Attention, the aggregating information from the i's L-hop temporal neighborhood as described in \cite{TGN}.

It's important to note that we searched for other nodes that already have kind of history, which means that already have memory features that are not initialized i.e, equal to zeros.

For simplicity, we decided to use $K=1$, only the most similar node.

\subsection{Define events-time difference}
For each node in the current batch, we calculate $\Delta T_{th}$, the events-time difference between the node's last updated event's timestamp and the current event's timestamp.

Using the $\alpha =0.025$ fraction of those events-time difference of the nodes in the current batch, we calculated the threshold value for $\Delta T$, as in equation \ref{T_th} For $ p = 1-\alpha = 0.975$.\\
It means we look for the relative staled nodes in the current batch. When we use the quantile we make sure that the threshold is conditioned with the current batch and we define staleness as relative to the other nodes, or users:

\begin{equation}
\resizebox{.6\hsize}{!}{
$\Delta T_{th} = Q(p) = \inf\{t \in \Re: p(t) \}$
}
\label{T_th}
\end{equation}

Thus, only nodes that their last event occurred before 
$\Delta T_{th}$ or longer will be updated with the similar nodes information. 
\section{Similarity Matrices}
\begin{figure}
    \centering
    \resizebox{0.95\columnwidth}{!}{\includegraphics{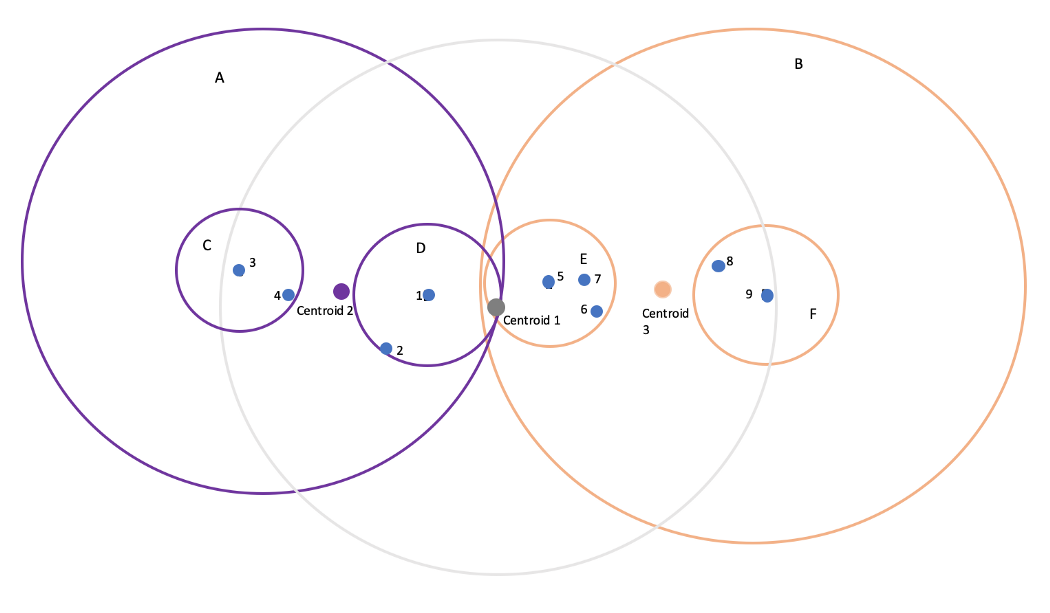}}
    \caption{Visualization of the Ball Tree Algorithm, \cite{TreeScience}}
    \label{fig:ball_tree}
\end{figure}

In order to find the most similar nodes with test to matrices of K-Nearest-Neighbors (KNN), using the 'sklearn' python package.
First, we used the KNN ball-tree algorithm, for fast generalized N-point problems. 
The ball-tree algorithm divides the data points into two clusters. Then, the process of dividing the data points into two clusters is repeated within each cluster until a defined depth is reached \cite{TreeScience}, as shown in ``Fig.~\ref{fig:ball_tree}.'' 

The second KNN algorithm we used is the brute-force search.

\section{Results and Discussion}
For two different similarity matrices and the TGN method as a baseline, we carried out three experiments. Our experiments were performed using two datasets: Wikipedia and Reddit. 

The results are drawn from the test set, which includes new nodes that the network was not exposed to during training or validation. 
We also checked the impact of the time-quantile threshold on the network performance, by decreasing the qunatile $q = 1 - \alpha$
for applying our similarity method on larger amount of nodes.

Table [\ref{tab:results_q_975}] shows the networks performance while using a fixed quantile of $q = 1 - \alpha = 0.975$, 
and Table [\ref{tab:diff_q}] shows the networks performance with different quantiles $q = {0.975, 0.80, 0.70}$. 




\begin{table}[htbp]
    \centering
    \begin{tabular}{|c|c|c|c|}
    \hline
    \cline{1-4} 
    model & dataset & AUC & precision  \\ \hline
    \textbf{TGN}  &  \textbf{wikipedia} & \textbf{0.963} & \textbf{0.967} \\ \hline
    Ball-Tree  &  wikipedia & 0.961 & 0.966 \\ \hline
    \textbf{Brute-force} &  \textbf{wikipedia} & \textbf{0.963} & \textbf{0.967} \\ \hline
    \\ \hline
    TGN  &  reddit & 0.953 & 0.958 \\ \hline
    \textbf{Ball-Tree}  &  \textbf{reddit} &  \textbf{0.957} & \textbf{0.960} \\ \hline
    Brute-force &  reddit & 0.953 & 0.957 \\ \hline

    \end{tabular}
    \caption{area under curve (AUC) and precision results for quantile $1 - \alpha = 0.975$}
    \label{tab:results_q_975}
\end{table}

\begin{table}[htbp]
    \centering
    \begin{tabular}{|c|c|c|c|}
    \hline
    \cline{1-4} 
    model & quantile & AUC & precision  \\ \hline
    TGN       &  -      & 0.963 & 0.967 \\ \hline
    Ball-Tree &  0.975  & 0.961 & 0.966 \\ \hline
    \textbf{Ball-Tree} &  \textbf{0.8}    & \textbf{0.966} & \textbf{0.970} \\ \hline
    Ball-Tree &  0.7    & 0.958 & 0.963 \\ \hline
    \end{tabular}
    \caption{different time quantile thresholds on Wikipedia dataset.}
    \label{tab:diff_q}
\end{table}

Model weights are available on request.


\section{Conclusion and Future-Work}
In the work, we study the influence of information aggregation from similar nodes on stale nodes. temporal deactivation of a node might effect the quality of recommendation system and edge prediction. 

Nodes may share similar or mutual information, so it can be exploit to express similarity when node becomes relative stale.  Though our solution has not showed significant improvement, there are future steps that may mitigate this issue by better effect on the stale nodes. 

Using a larger number of similar nodes may increase their effectiveness and improve network performance. 
Also, decreasing the time-quantile threshold may increase the impact of our new method, by apply it on larger number of nodes. 
In addition, applying attention mechanism or learnable nodes similarity network to find the most similar combination of nodes instead of KNN also may improve the staleness performance.
Moreover, there is future work on an evaluation metric that will be specific for the stale nodes.

\bibliographystyle{IEEEtran}
\bibliography{ref}

\end{document}